\documentclass[a4paper,11pt]{article}
\usepackage{pos}
\usepackage{physics,cancel}

\title{Discretisation effects of gradient flows in QCD-like theories on the lattice}

\author*[a]{Pietro Butti}
\author[a,b]{Michele Della Morte}
\author[a,b,c]{Benjamin J\"{a}ger}
\author[d]{Sofie Martins}
\author[e]{J.~Tobias Tsang}

\affiliation[a]{Dept. of Mathematics and Computer Science, University of Southern Denmark, 5230 Odense M, Denmark}

\affiliation[b]{Quantum Theory Center ($\hbar$QTC), University of Southern Denmark, 5230 Odense M, Denmark}
\affiliation[c]{Danish Institute for Advanced Study, University of Southern Denmark, 5230 Odense M, Denmark}
\affiliation[d]{Institute of Physics, NAWI Graz, University of Graz, 
   8010 Graz, Austria}
\affiliation[e]{Department of Mathematical Sciences, University of Liverpool, Liverpool L69 3BX, United Kingdom}

\emailAdd{pbutti@qtc.sdu.dk}

\abstract{
Recent software advances now allow large-scale lattice studies of the Corrigan--Ramond large-$N_C$ limit of Yang-Mills theory coupled with a two-index antisymmetric fermion, providing a path to SUSY Yang-Mills. We are currently generating ensembles for $N_C=4,5,6$ for lattice spacings in the range $0.11 - 0.08$ fm.
We report on two aspects of our work: the study of topological properties as well as estimates of discretisation effects.
The first aspect is relevant since naively, fractional topological charges might be expected in our simulations. Using a gluonic definition of the topological charge combined with gradient flow, we perform an analysis of the effect of different discretisations of the kernel action, from which we identify and interpret quantitative differences between Wilson and over-improved flows such as DBW2.
The second aspect is addressed by considering ratios of different reference flow times. We conclude that our current simulations might be affected by discretisation effects of order 10\%. 

}

\FullConference{The 42nd International Symposium on Lattice Field Theory (LATTICE2025)\\
2-8 November 2025\\
Tata Institute of Fundamental Research, Mumbai, India\\}

\begin{document}
\maketitle

\section{Introduction}
We report developments in our recent efforts to numerically simulate orientifold theories, namely one-flavour vector gauge theories with a single Dirac fermion in the two-index anti-symmetric representation of a SU($N_C$) gauge group.
In the large-$N_C$ limit orientifold theories are equivalent to super-Yang-Mills~\cite{Armoni:2003gp,Armoni:2003fb} and this equivalence has been exploited to obtain a number of predictions including the value of the gluino condensate~\cite{Armoni:2003yv}, which
we plan to compute in the future through the Banks-Casher relation. 
That is one of the reasons why  we started exploring methods to measure the topological charge, as we discuss here.
The planar equivalence mentioned above has further been used to derive effective low-energy 
Lagrangians~\cite{Sannino:2003xe}, in order to describe the spectrum and the vacuum properties of orientifold theories (see Ref.~\cite{Sannino:2024xwj} for a recent review). The long-term plan of this effort is to test such
effective theories through non-perturbative computations of the spectrum by means of lattice simulations and determine the relevant low-energy constants. 
Previous reports in this direction appeared in Refs.~\cite{Ziegler:2021nbl,Jaeger:2022ypq,DellaMorte:2023ylq,Martins:2023kcj,DellaMorte:2025tks,Butti:2025rnu}. In the following we briefly summarise the most recent results and discuss the progress in considering $N_C=4, 5$ and $6$ as well as in addressing discretisation effects.  
Such developments define the state of the art in lattice simulations of orientifold theories.
The remainder of this work is organised as follows: in Section~\ref{sec:flows} we summarise our recent results from Ref.~\cite{Butti:2025rnu}. 
A similar study for $SU(2)$ pure gauge theory has been performed in Ref.~\cite{Tanizaki:2024zsu}.
In Section~\ref{sec:artifacts} we take a first look at discretisation effects before providing an outlook in Section~\ref{sec:conc}.

\section{Exploring different flows\label{sec:flows}}
The topological charge is known to suffer from critical slowing down as the lattice spacing is reduced and as $N_C$ increases~\cite{Schaefer:2010hu}. 
Therefore, it serves both as a physical quantity of interest and as a diagnostic for algorithmic performance. Beyond algorithmic considerations, the behaviour of the topological charge is also of conceptual interest in these theories. For fermions in the two-index anti-symmetric representation, group-theoretical arguments suggest that the winding number may be quantised in units of $1/(N_C-2)$~\cite{Leutwyler:1992yt,Fodor:2009nh}. However, it is known that with periodic boundary conditions (pBC) and sufficiently smooth gauge fields, the topological charge is expected to be integer-valued in the continuum limit. Fractional topological charges are instead known to persist when twisted boundary conditions are used~\cite{Gonzalez-Arroyo:2019wpu,GarciaPerez:2000aiw}. Since our setup uses pBCs, possible observations of non-integer values should therefore be interpreted as lattice artefacts rather than genuine continuum features.
We performed a quantitative analysis of such a statement in Ref.~\cite{Butti:2025rnu}, of which we provide a summary in this section. The reader is invited to refer to Ref.~\cite{Butti:2025rnu} for more details.

\subsection{Numerical setup}
We generated an auxiliary set of coarser ensembles, summarised in Tab.~\ref{tab:ens},
\begin{table*}
\centering
\begin{tabular}{l|ccccccccc}
     name & $N_C$ & $\beta$ & $\kappa$ & $L/a$ & $t_0/a^2$  & $L/\sqrt{t_0}$ & $M_\pi^\mathrm{conn.} \sqrt{t_0}$ & $N_\mathrm{conf}$
     \\\hline\hline
     N4L12  & 4&  7.1 & 0.15770 & 12 & 0.8691(12) &12.8720(90)& 0.6042(7)  & 648\\
     N4L14  & 4&  7.2 & 0.15651 & 14 & 1.3641(17) &11.9868(75)& 0.5958(7) & 677\\
     N4L16  & 4&  7.3 & 0.15525 & 16 & 1.8592(19) &11.7344(60)& 0.5931(6) & 640\\
     N4L18  & 4&  7.4 & 0.15401 & 18 & 2.3885(21) &11.6470(51)& 0.6103(5) & 583\\  
     \hline
     N5L12 & 5& 11.3 & 0.16089 & 12 & 1.1171(11) &11.3536(55)& 0.8002(5) & 1081\\
     N5L14 & 5& 11.4 & 0.16026 & 14 & 1.4800(14) &11.5078(53)& 0.8058(5) & 704\\
     N5L16a & 5& 11.5 & 0.15959 & 16 & 1.8422(15) &11.7884(49)& 0.7575(5) & 565\\
     N5L16b &5& 11.55& 0.15929 & 16 & 2.0409(21) &11.1997(58)& 0.8460(5) & 469\\\hline
     N6L16 & 6& 16.5 & 0.16200 & 16 & 1.33944(64)&13.8248(33)& 0.9668(3) & 742\\
\end{tabular}
\caption{Parameters of the ensembles used in this study. The temporal extent is fixed to be $T = 3 L$ for all ensembles. Quoted uncertainties are statistical only. Ensembles used in Ref.~\cite{Butti:2025rnu}.}
\label{tab:ens}
\end{table*}
in order to study the discretisation effects on the topological charge. We use a gluonic definition evaluated on smoothened gauge fields and systematically vary both the flow time and the discretisation of the flow equation. In particular, we are interested in studying a topological charge definition which allows for the identification of eventual fractional values and their behaviour in the continuum limit.

To deal with UV-fluctuations of the topological charge, we smeared our configurations according to the standard procedure of \textit{gradient flow}~\cite{Luscher:2010iy}. On the lattice, the flow equation must be specified through a choice of kernel action. Different choices correspond to different lattice artefacts at finite flow time, while all definitions become equivalent in the continuum limit. We define the flow using a gauge action composed of plaquette and rectangular Wilson loops,
\begin{equation}
    S_{\text{flow}} = c_0 S_{\text{plaq}} + c_1 S_{\text{rect}},\quad c_0 + 8c_1 = 1\,,
\end{equation}
where the standard choice, known as the Wilson flow, corresponds to  $c_1=0$. In this case the flow is generated solely by the plaquette action and represents the simplest discretisation of the continuum gradient flow. In addition to the Wilson flow, following the prescription in Ref.~\cite{GarciaPerez:1993lic} we consider an \textit{over-improved} flow, obtained by a particular choice of $c_1$. In particular, we focus on the DBW2 kernel, characterised by $c_1 = -1.4088$~\cite{QCD-TARO:1996lyt,QCD-TARO:1999mox}. 
The term ``over-improvement'' refers to a choice of $c_1$ leading to a positive sign of the $\mathcal O(a^2)$ correction term in the classical instanton action. 
Contrary, for under-improved flows such as the Wilson flow, the correction is negative and small-size instantons can induce transitions between topological sectors and lead to an unstable definition of the topological charge at finite lattice spacing.
Regardless of the definition of the kernel action, the flow time $t$ at which $Q$ is computed should always be chosen in the window
\begin{equation}
    2a \ll \sqrt{8t} \ll \frac{L}{2}\,,
\end{equation}
to mitigate large discretisation and finite-volume effects, respectively.

\subsection{Results and discussion}
\begin{figure}
    \centering
    \includegraphics[width=0.5\linewidth]{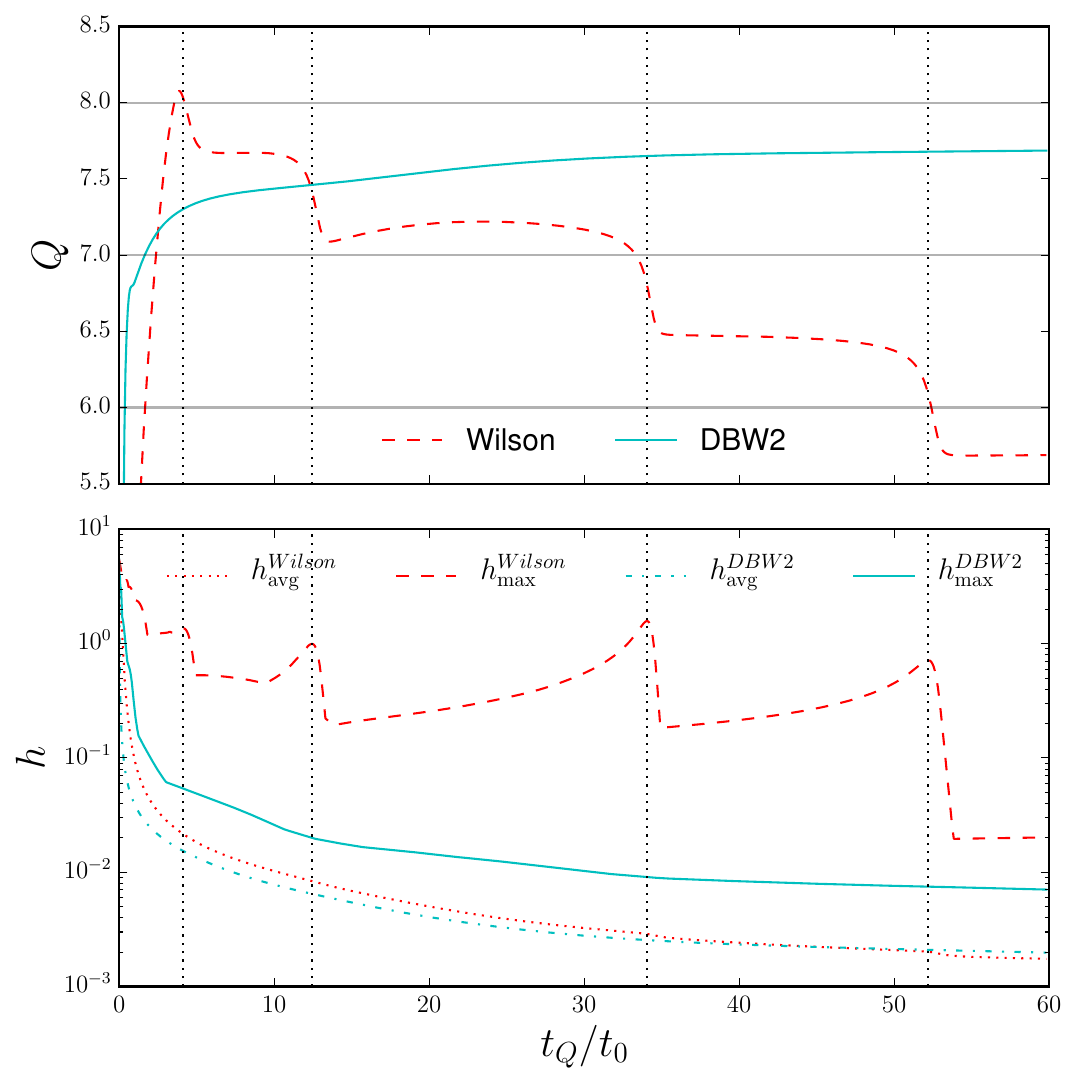}
    \caption{Topological charge (top) as well as $h_\text{avg}$ and $h_\text{max}$ (bottom) for the Wilson flow (red, dashed and dotted) and the DBW2 flow (cyan, solid and dash-dotted) as a function of very long flow times and on a single configuration of the N4L12 ensemble. The vertical dotted lines correspond to the location of local maxima in $h^\text{Wilson}_\text{max}$ which correlate with transitions of $Q^\mathrm{Wilson}$. Figure taken from Ref.~\cite{Butti:2025rnu}.}
    \label{fig:hflow}
\end{figure}

We determine the topological charge using a gluonic definition evaluated on gauge fields smoothened by the gradient flow. For each configuration, the charge $Q$ is computed from a clover discretisation of the field-strength tensor constructed from the flowed links,
\begin{equation}
    Q = \frac{1}{32\pi^2}\sum_x \epsilon_{\mu\nu\rho\sigma} \mathrm{tr}[\hat C_{\mu\nu}(x)\hat C_{\rho\sigma}(x)]\,,
\end{equation}
where $\hat C$ is the clover discretisation of the Yang-Mills field strength. The behaviour of $Q$ is then monitored as a function of the flow time $t$ for both the Wilson and the DBW2 flow. 
\begin{figure}
    \centering
    \includegraphics[width=0.35\linewidth]{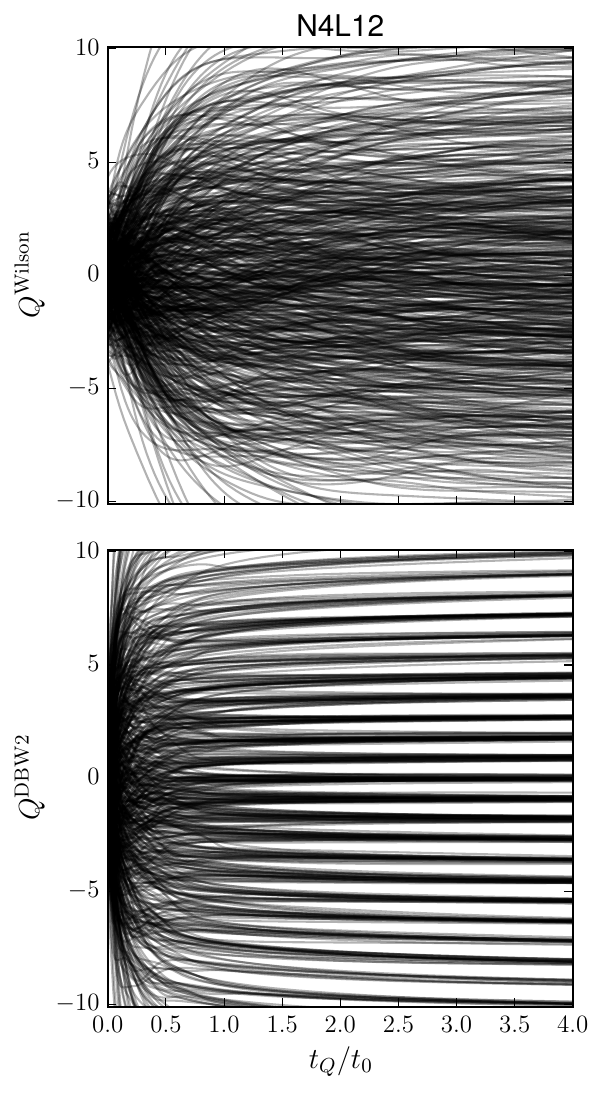}
    \includegraphics[width=0.35\linewidth]{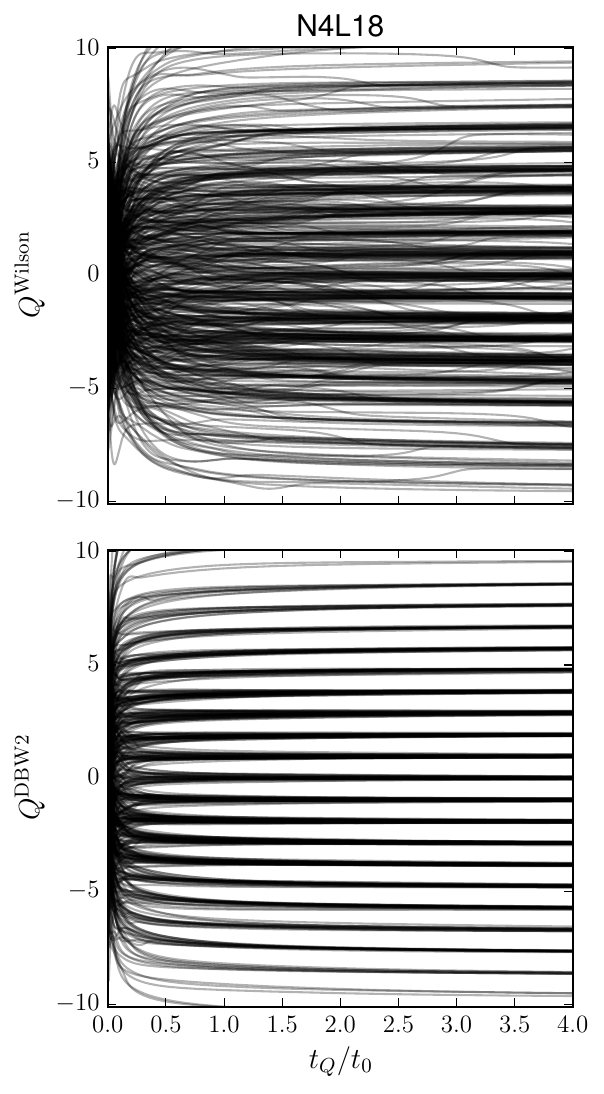}
    \caption{Topological charge determined using the Wilson flow (top) and the DBW2 flow (bottom) on the coarsest (left) and finest (right) $N_C=4$ ensembles. Figure taken from Ref.~\cite{Butti:2025rnu}.}
    \label{fig:flowQ}
\end{figure}
In Fig.~\ref{fig:flowQ}, we find that the DBW2 flow typically produces well-defined plateaux at significantly smaller flow times than the Wilson flow. In contrast, under the Wilson flow, the charge often undergoes discrete jumps at intermediate and even relatively large flow times, indicating topology-changing events induced by lattice artefacts. The frequency of such jumps decreases as the continuum limit is taken.
To investigate the nature of such jumps, we computed the flow time dependence of a local smoothness observable, constructed from plaquette values $p$ as
\begin{equation}
    h(p) = \mathrm{Re\,}\mathrm{tr}\,\bigg[1 - \prod_{(x,\mu)\in p} U_{x,\mu}(t)\bigg]
\end{equation}
and distinguishing between $h_\text{max}=\max_{p}h(p)$ and $h_\text{avg}=\mathrm{avg}_{p}h(p)$. We observe that $h^\mathrm{Wilson}_\mathrm{avg}$ monotonically decreases as the smoothening flow proceeds, whilst $h_\mathrm{max}^\mathrm{Wilson}$ displays local maxima, as is illustrated in Fig.~\ref{fig:hflow}. We further find, that these maxima correlate with transitions of the topological charge $Q^\mathrm{Wilson}$. We interpret these ``spikes'' in $h_\mathrm{max}^\mathrm{Wilson}$ as an indicator of the presence of local structures that can mediate changes in the topological sector during the flow. 
This is compatible with the observation that the Wilson flow tends to enhance the effect of small instantons which produces instabilities of the topological charge. In the DBW2 case, both $h_\text{avg}$ and $h_\text{max}$ are monotonously decreasing in time.

Based on these observations, we define the topological charge by selecting a reference DBW2-flow time $t_Q$ within the plateau region in $t$ and consider the corresponding histogram of $Q(t_Q)$. As argued in Ref.~\cite{Butti:2025rnu} we observe that the distribution of $Q$ is clustered around equally spaced pseudo-integers. With an opportune choice of $t_Q$, the occurrence of a configuration with a fractional value of $Q$ completely disappears. Nevertheless, even with the most conservative choice for $t_Q$, the bins corresponding to fractional $Q$ get rapidly depleted as $a\rightarrow 0$.

\begin{table*}
\centering
\begin{tabular}{ccccccccc}
     $N_C$ & $\beta$ & $\kappa$ & $L/a$ & $t_0/a^2$  & $L/\sqrt{t_0}$ & $M_\pi^\mathrm{conn.} \sqrt{t_0}$ & $N_\mathrm{conf}$
     \\\hline\hline

3 & 4.17 & 0.14268 & 12 & 2.47 & 7.64 & 0.26 & 1318 \\
3 & 4.17 & 0.14319 & 14 & 2.53 & 8.80 & 0.22 & 1053 \\
3 & 4.17 & 0.14268 & 16 & 2.46 & 10.20 & 0.26 & 1705 \\
3 & 4.17 & 0.14370 & 16 & 2.63 & 9.87 & 0.18 & 1082 \\
3 & 4.17 & 0.14255 & 18 & 2.44 & 11.52 & 0.27 & 1640 \\
{\bf 3} & {\bf 4.17} & {\bf 0.14255} & {\bf 18} & {\bf 2.69} & {\bf 10.97} & {\bf 0.15} & {\bf 1022} \\
3 & 4.17 & 0.14422 & 20 & 2.73 & 12.10 & 0.12 & 1034 \\
\hline
4 & 7.5 & 0.15258 & 12 & 2.90 & 7.05 & 0.24 & 930 \\
4 & 7.5 & 0.15295 & 14 & 3.02 & 8.06 & 0.21 & 700 \\
4 & 7.5 & 0.15258 & 16 & 2.90 & 9.40 & 0.24 & 774 \\
4 & 7.5 & 0.15328 & 16 & 3.13 & 9.04 & 0.18 & 680 \\
4 & 7.5 & 0.15343 & 18 & 3.19 & 10.08 & 0.16 & 802 \\
{\bf 4} & {\bf 7.5} & {\bf 0.15356} & {\bf 20} & {\bf 3.23} & {\bf 11.13} & {\bf 0.15} & {\bf 572} \\
4 & 7.5 & 0.15379 & 24 & 3.34 & 13.13 & 0.12 & 720 \\
\hline
5 & 11.6 & 0.16000 & 12 & 2.54 & 7.53 & 0.25 & 946 \\
5 & 11.6 & 0.15258 & 14 & 2.64 & 8.60 & 0.21 & 414 \\
5 & 11.6 & 0.16030 & 16 & 2.76 & 9.63 & 0.18 & 612 \\
5 & 11.6 & 0.16070 & 18 & 2.82 & 10.72 & 0.16 & 449 \\
{\bf 5} & {\bf 11.6} & {\bf 0.16080} & {\bf 20} & {\bf 2.89} & {\bf 11.76} & {\bf 0.15} & {\bf 232} \\
\hline
6 & 16.5 & 0.16566 & 12 & 2.11 & 8.26 & 0.27 & 457 \\
6 & 16.5 & 0.16593 & 14 & 2.21 & 9.42 & 0.23 & 121 \\
6 & 16.5 & 0.16617 & 16 & 2.32 & 10.50 & 0.20 & 66 \\
6 & 16.5 & 0.16627 & 18 & 2.39 & 11.64 & 0.18 & 113 \\
{\bf 6} & {\bf 16.5} & {\bf 0.16637} & {\bf 20} & {\bf 2.46} & {\bf 12.75} & {\bf 0.15} & {\bf 58} \\
     \hline
   
\end{tabular}
\caption{Parameters of the ensembles used to explore the $N_C$ dependence. The temporal extent is fixed to be $T/a = 48$ for all ensembles. Each configuration is separated by 4 trajectories with 2 MDU units.}
\label{tab:current_ensembles}
\end{table*}

\section{A first look at lattice artefacts \label{sec:artifacts}}
In this section we make use of the gradient flow data we generated to discuss the impact of lattice artefacts on scale setting. 
We start by concentrating on the subset of coarser ensembles we generated to investigate the topological smoothing properties of the gradient flow.
To set the lattice scale we integrate the gradient flow and compute the expectation value of the flowed energy density $\langle E(x,t)\rangle = \langle \mathrm{tr} \left(G_{\mu\nu}(x,t)G^{\mu\nu}(x,t)\right) \rangle$, choosing the plaquette discretisation for the action density. The relative lattice scale is usually set by determining the value of flow time $t_c$ for which $\langle t^2 E_\text{plaq}\rangle$ takes a particular value. For the case of $N_C = 3$ this value is conventionally chosen to be $c=0.3$. Generalising this for $N_C > 3$ whilst accounting for the leading $N_C$ dependence of $\langle E \rangle$, and hence staying a fixed 't Hooft coupling, requires a rescaling by $(N_C^2-1)/N_C$ and so yields
\begin{equation}\label{eq:t0rescaling}
    \langle t_c^2 E_\text{plaq}\rangle = c\qty(\frac{N_C^2-1}{N_C})\frac{3}{8}\,. 
\end{equation}
Different reference values $c$, lead to a different definitions of the associated flow scale $t_c$. In this work we will consider canonical choice $c=0.3$ as well as $c=0.5$ and (following the conventional notation) call the corresponding flow times $t_0$ and $t_1$.
We use the dimensionless ratio $R=t_0/t_1$ to explore the effect of the lattice
discretisation, similar to the procedure in Ref.~\cite{RamosMartinez:2023tvx}.

\begin{figure}
    \centering
    \includegraphics[width=1\linewidth]{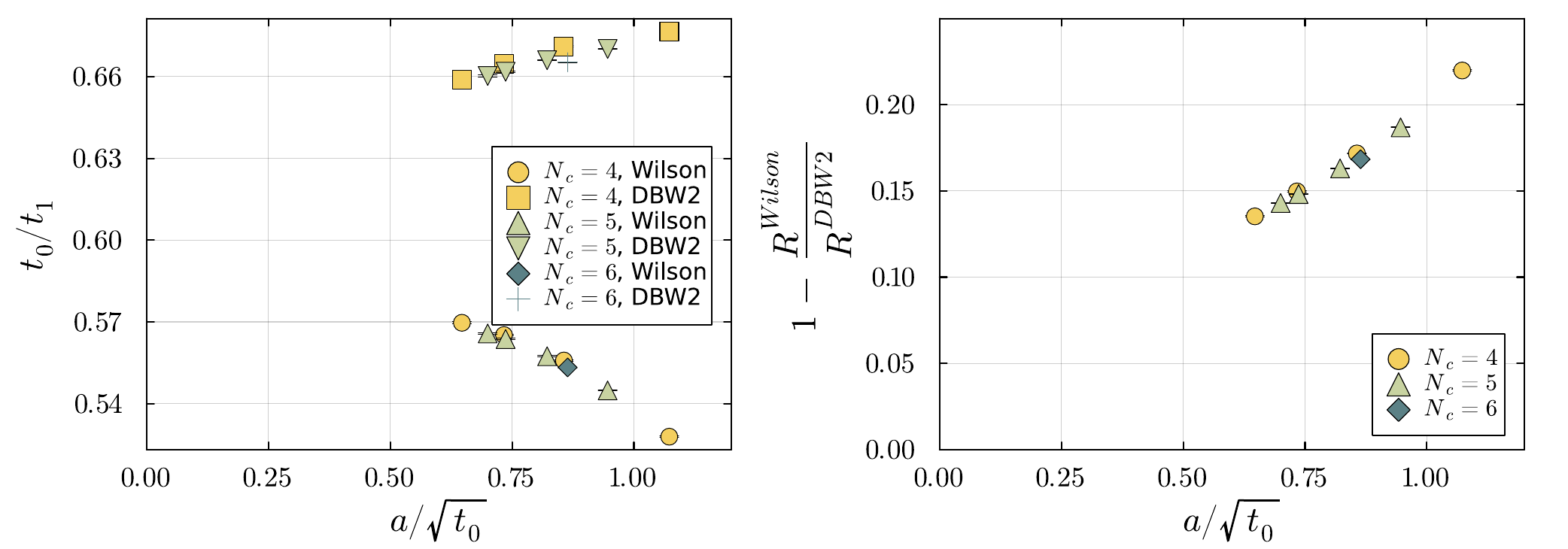}
    \caption{\textit{Left-hand side panel:} ratio of scales $R=t_0/t_1$ vs lattice spacing in units of $\sqrt{t_0^{\text{Wilson}}}$ for the ensembles in Tab.~\ref{tab:ens} for different kernel action in the gradient flow discretisation. \textit{Right-hand side panel:} difference from unity of the ratio $R^{\text{Wilson}}/R^{\text{DBW2}}$ for the same ensembles vs $a/\sqrt{t_0}$. 
    }
    \label{fig:topo_art}
\end{figure}

In Fig.~\ref{fig:topo_art}, we plot the values of the ratio of scales as a function of the lattice spacing (in units of $t_0^\text{Wilson}$) for the different auxiliary ensembles we generated for Wilson and the DBW2 flow scales.
First, we notice that results for different values of $N_C$ align nicely with each other at finite lattice spacing, which indicates that the perturbative rescaling performed in Eq.~\eqref{eq:t0rescaling} works very well.
Second, we observe that the discretisation effects associated with DBW2 flow contribute with opposite sign with respect to the Wilson flow case. 
Concerning the magnitude of lattice artefacts, although the effect on DBW2 seems somewhat milder, we conclude that they are both of the order of 10\%. In the right-hand panel of Fig.~\eqref{fig:topo_art} we also plot $1-R^\text{Wilson}/R^\text{DBW2}$, which is expected to vanish in the continuum limit. We attempted a constrained extrapolation to $a=0$ in units of $t_0$ through a sum of cubic and  quadratic terms in $a/\sqrt{t_0}$, which confirms the expected behaviour in the continuum limit~\cite{Ramos:2015baa}. A more sophisticated extrapolation is out of the scope of this work, as the coarse values of $a$ combined with the per-cent precision of our scales influence the quality of the fit.

We recall that the difference between DBW2 and Wilson lattice artefacts concerns mainly the discretisation effects associated with the smoothening action of the gradient flow~\cite{Ramos:2015baa}. Nevertheless, by only looking at the Wilson flow scales, one can get a rough idea of the order of magnitude of the artefacts we might expect in our future studies. In this regard, we performed the same scale setting analysis on some of the ensembles we plan to use for the spectroscopy study. We selected one ensemble for each $N_C$ value by keeping the \emph{fake-pion} mass\footnote{We define the fake pion through the connected piece of the pseudoscalar correlator.} approximately fixed at $\sim 520(20)$ MeV (converting to physical units from the Wilson flow scale using the standard reference value for QCD for comparison).
\begin{figure}
    \centering
    \includegraphics[width=0.8\linewidth]{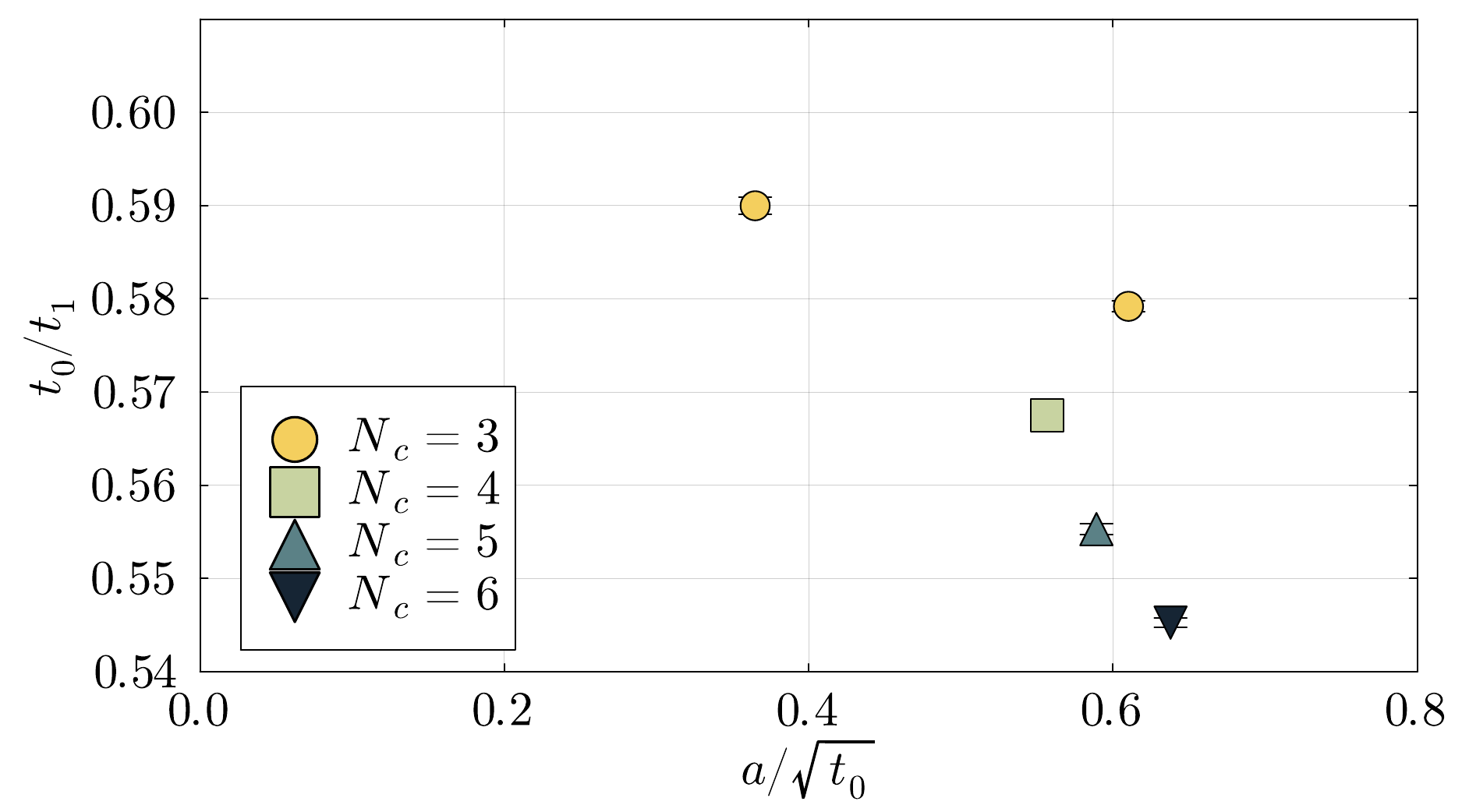}
    \caption{Ratio of scales $t_0/t_1$ for representative ensembles marked in bold in Tab.~\ref{tab:current_ensembles}. The leftmost $N_C=3$ point (yellow circle corresponding to $a/\sqrt{t_0}\simeq 0.39$) has been taken from Ref.~\cite{DellaMorte:2023ylq}.}
    \label{fig:scales_prod_art}
\end{figure}
In Fig.~\ref{fig:scales_prod_art}, we plot the value of $t_0/t_1$  for these ensembles obtained through the Wilson flow in the plaquette discretisation. The different values of $a/\sqrt{t_0}$ are mostly due to a mismatch in volume and in fake-pion mass. However, we observe that the magnitude of the discretisation effect on the Wilson flow scale is also of order 10\%.


\section{Conclusions and Outlook \label{sec:conc}}
{Orientifold theories can now be efficiently simulated and software packages which can be used on different architectures are publicly available~\cite{Drach:2025eyg}.
This allows to non-perturbatively check the conjectured equivalence
with the mesonic sector of super-Yang-Mills in the large $N_C$ limit.
We reported our progress in that direction. In particular we discussed the ensembles generation for $N_C=4,5,6$, extending our previous results in Ref.~\cite{DellaMorte:2023ylq} for $N_C=3$. 
We also presented a first attempt at assessing cutoff effects from lattice spacings of about $0.1$ fm with a tree-level improved lattice action.
In general, the precision requirement for this type of study is lower than typical QCD calculations,  however since cutoff effects of the order of $10\%$ can be expected 
and will be necessary to perform a continuum limit when aiming at a few-percent accuracy.
In this spirit, we have now added a second lattice resolution for $N_C=3$ and plan to have the same for most of the other $N_C$ values. That would lead to the most solid lattice study of such theories, with reliable estimates of all systematic effects.
}

\section*{Acknowledgements}
The authors thank Nikolai Husung for useful comments.
We performed parameter tuning on the Discoverer supercomputer (grant EHPC-REG-2023R01-102) and generated configurations on the LUMI supercomputer (allocation EHPC-EXT-2024E01-038) provided by EuroHPC JU. This work was further supported by a DeiC grant (case 4265-00016A) and the UCloud DeiC Interactive HPC system, managed by the eScience Center at the University of Southern Denmark. S.M. has received funding from the Austrian Science Fund FWF under grant number PAT6443923.

\bibliographystyle{apsrev4-1}
\bibliography{biblio}

\end{document}